\documentclass[11pt]{article}
\usepackage{amsmath,amsfonts, amssymb,graphicx,geometry,authblk,color,soul,comment,hyperref,amsthm,wrapfig,xcolor} 
\usepackage{hhline}
\usepackage{mathtools}

\usepackage{subcaption}
\usepackage[italicdiff]{physics}
\title{{Learning constitutive relations from experiments: \\ II. Dynamic indentation}}
\author[1]{Andrew Akerson}
\author[2]{Aakila Rajan}
\author[3]{Daniel Casem}
\author[2]{Kaushik Bhattacharya\footnote{Corresponding Author: bhatta@caltech.edu}}
\affil[1]{{Massachusetts Institute of Technology, Department of Mechanical Engineering, Cambridge, MA 0213}}
\affil[2]{Division of Engineering and Applied Science, California Institute of Technology, Pasadena, CA 91125, USA}
\affil[3]{CCDC Army Research Laboratory, Aberdeen Proving Ground, Aberdeen, Maryland 21005, USA}
\date{}

\begin{document}
\maketitle

\begin{abstract}
We continue the development of a method to accurately and efficiently identify the constitutive behavior of complex materials through {experimental} observations that we started in \cite{akerson2024learning}.   We formulate the problem of inferring constitutive relations from experiments as an indirect inverse problem that is constrained by the balance laws. Specifically, we seek to find a constitutive behavior that minimizes the difference between the experimental observation and the corresponding quantities computed with the model, while enforcing the balance laws. We formulate the forward problem as a boundary value problem corresponding to the experiment, and compute the sensitivity of the objective with respect to the model using the adjoint method.  In this paper, we extend the approach to include contact and study dynamic indentation.  Contact is a nonholonomic constraint, and we introduce a Lagrange multiplier and a slack variable to address it.  We demonstrate the method on synthetic data before applying it to experimental observations on rolled homogeneous armor steel and a polycrystalline aluminum alloy.

\end{abstract}

\section{Introduction}
The engineering of solutions involving complex materials and phenomena requires a constitutive relation that describes the properties of the material.  This constitutive relation is typically obtained empirically by conducting experiments. {However, we cannot directly deduce the constitutive relation from measurements, and it must be obtained through the} solution of an inverse problem.   In fact, we cannot even directly measure the quantities like stress, strain, strain rate, and energy density that comprise the constitutive relation.  Instead, these must be inferred from quantities like displacements and total forces that can be measured in the laboratory.  Thus, the problem of inferring constitutive relations from experiments, and thereby completing the continuum mechanical formulation, requires the solution of an inverse problem.

In part I of this work~\cite{akerson2024learning}, we propose an approach where this indirect inverse problem is formulated as a partial differential equation (PDE) constrained optimization problem.  { The constitutive law is postulated as a parametrized relation.}  The problem of identifying the constitutive relation, then, is to find the parameters that minimize the difference between the experimental observation and the corresponding quantities computed by solving the governing equations (PDE constraint). {This is a similar to finite element method updating (FEMU) schemes~\cite{BermanNagy1983}.  However, we adopt a gradient-based optimization approach in which we compute the sensitivity using the adjoint method. Recently, researchers have independently developed differentiable methods for elasto-plastic characterization~\cite{KUMAR2025,FERREIRA2026} through automatic differentiation and adjoint-based approaches. For a further discussion of the literature, we direct the reader to part I of the work~\cite{akerson2024learning}, where} we survey the literature, describe the method broadly and demonstrate it against quasistatic and dynamic experiments using synthetic data.

In this part II, we extend the work to dynamic indentation, and demonstrate the method with both synthetic and experimental data.   Since their introduction by J.A.\ Brinnell over a century ago, indentation tests have been widely used in the static setting to characterize the hardness of materials.  Koeppel and Subhash~\cite{koeppel_1997} introduced dynamic indentation by adapting a split Hopkinson pressure bar to study the dynamic hardness of materials.  Since then, this method, and adaptations based on it, have been used by various researchers due to the relative ease of execution~\cite{anton_2000,nilsson,si_dynamic_2023,casem2023kolsky}.  However, obtaining quantitative information about the constitutive behavior has remained a challenge for a number of reasons (e.g.,~\cite{lee_dynamic_2018}).  Dynamic indentation leads to a complex, time-dependent, and heterogeneous state of stress with elastic-plastic deformation.  It thus enables probing multiple states at once.  However, obtaining this quantitative information from this complex field requires the solution to a challenging inverse problem.  Fortunately, this inverse problem is ideally suited for the proposed approach.

Indentation, however, involves contact between the indenter and the material being probed.  Contact is a nonholonomic, one-sided constraint, and this requires additional theoretical development to the formulation presented in part I.  This is one of the main objectives of this paper. The unilateral contact constraint has been addressed in the literature through a variety of methods including penalty-based~\cite{laursen2003computational}, classical Lagrange multiplier~\cite{wriggers2006computational,kikuchi_contact_1988,Huber2008}, augmented Lagrangian~\cite{simo1992}, third-medium~\cite{wriggers2013}, or {incremental potential contact methods~\cite{Li2020}. The incremental potential method relaxes the zero-separation contact constraint resulting in a smooth contact force, and this has been exploited to conduct design optimization through an adjoint approach~\cite{Huang2024}. For this work, we take a different approach. We strictly} enforce the contact condition in the forward problem through a Lagrange multiplier and a slack variable, and carry out the computations through an efficient staggered numerical scheme~\cite{CAMACHO1997269}. We derive the adjoint equations and apply a similar computational algorithm to its evolution.  We limit ourselves to frictionless contact, but note that we can treat friction similarly.

The second objective of this work is to demonstrate the method against experimental data.  We do so using experimental data taken on both rolled hardened armor (RHA) steel, and an aluminum alloy.  In each case, we demonstrate the ability of our method to recover elasto-plastic constitutive behavior using only a few tests.

{ In parts I and II of the work, we postulate the constitutive relation as a parametrized relation and use the proposed method for identifying the parameters.  In the forthcoming part III, we extend the work to identify the constitutive relation postulated in the form of a neural network.  The identified neural network can then be used to identify the functional form of the constitutive relation.}

This paper is divided into 5 sections. We begin in Section~\ref{sec:form} by describing the formulation and the method.  In Section~\ref{sec:gen_form} we present the forward problem associated with dynamic indentation, formulating this with the contact law.  We then derive the adjoint equations and the resulting algorithm for solving the inverse problem in Section~\ref{sec:opt_for_viscoplastic_param}.  We demonstrate the method using synthetic data in Section \ref{sec:ex}. Then, in Section~\ref{sec:expt_inv}, we apply the method to experimental observations. We conclude in Section~\ref{sec:conclusion}.

\section{Formulation and method} \label{sec:form}
\begin{figure}
    \centering
    \includegraphics[width=0.6\textwidth]{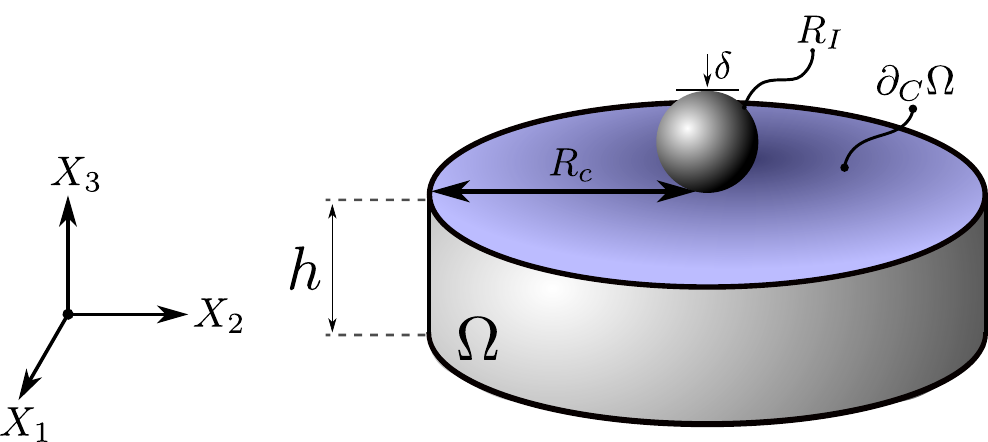}
    \caption{Schematic diagram of the rigid indentation test. We consider a cylindrical domain $\Omega$ of radius $R_C$ and height $h$ resting on a rigid surface indented on its top surface $\partial_C \Omega$ by a rigid sphere of radius $R_I$ to a prescribed depth $\delta$.}
    \label{fig:indent_diagram}
\end{figure}

The schematic diagram of a typical indentation test is shown in Figure~\ref{fig:indent_diagram}.
We consider an elastic-plastic body occupying the domain $\Omega \subset \mathbb{R}^N$ over time $(0, T)$.   We assume that part of the boundary of the body is subjected to a fixed displacement, $u = u_0$ on $\partial_u \Omega$ where $u : (0, T) \times \Omega \to \mathbb{R}^N$ denotes the displacement field and $u_0$ is given. The body is subjected to a body force per unit volume $b$. A rigid indenter occupying $\Omega_I \subset \mathbb{R}^N$ with centroid $x_I: (0,T) \to \mathbb{R}^N$ and orientation $o_I:  (0,T) \to $SO$(N)$ is brought in contact with a part of the boundary $\partial_C \Omega$, with the rest of the boundary being traction-free.  We measure the evolution of the reaction force on the indenter, and seek to use this information to find the constitutive behavior of the material.

\subsection{Governing equations} \label{sec:gen_form}
\paragraph{Constitutive Model} We assume that the body has density $\rho$, and is governed by small strain, J2 plasticity with power-law isotropic hardening and power law rate-sensitivity~\cite{Ortiz1999,lubliner2008plasticity}.  The elastic behavior is linear, and the stored energy density is $ \frac{1}{2} \varepsilon^e \cdot \mathbb{C} \varepsilon^e$ where  $\mathbb{C}$ is the stiffness tensor, $\varepsilon^e \coloneqq \varepsilon - \varepsilon^p$ is the elastic strain, $\varepsilon \coloneqq (\nabla u + \nabla u^T)/2$ is the total strain and $\varepsilon^p$ is the plastic strain. We consider isotropic strain hardening described by a power-law plastic potential
\begin{equation}
    W^p(q) = \sigma_y\left[ q + \frac{n \varepsilon^p_0}{n + 1}\left( \frac{q}{\varepsilon^p_0}\right)^{(n+1)/n}\right],
\end{equation}
where $q : (0, T) \times \Omega \to \mathbb{R}$ is the accumulated plastic strain defined by
\begin{equation}
    \dot{q} = \sqrt{\frac{2}{3} \dot{\varepsilon^p} \cdot \dot{\varepsilon^p}},
\end{equation}
$\sigma_y$ is the yield stress, $\varepsilon^p_0$ is the reference plastic strain, and $n$ is the hardening exponent. We introduce the rate-dependence through a dissipation potential
\begin{equation}
    \psi (\dot{q}) = \begin{cases} 
     g^*(\dot{q}) = \frac{ m \sigma_y \dot{\varepsilon}^p_0}{m + 1}\left( \frac{\dot{q}}{\dot{\varepsilon}^p_0}\right)^{(m + 1)/m} & \ \dot{q} \geq 0, \\
    \infty & \ \dot{q} < 0
    \end{cases},
\end{equation}
where $\dot{\varepsilon}^p_0$ is the reference plastic strain rate, and $m$ is the rate-sensitivity exponent. 

\paragraph{Dynamic Rigid Contact} \label{sec:forward_con}
We assume that the contact with the rigid body is frictionless for simplicity, but note that the formulation may be extended to include friction.  We describe the rigid body and its trajectory by a contact function $\mathcal{C}_I : \mathbb{R}^N \times \mathbb{R}^N \times \text{SO}(N) \to \mathbb{R}$ such that $\mathcal{C}_I(x, x_I,o_I) < 0$ if $x \in \Omega_I$ and $\mathcal{C}_I(x, x_I,o_I) \geq 0$ if $x \notin \Omega_I$. In the case of a spherical indenter of radius $R_I$, the contact function may take the form
\begin{equation}
    \mathcal{C}^{sphere}_I(x, x_I) = |x - x_I| - R_I.
\end{equation}
It does not depend on orientation due to symmetry. In this seting, frictionless contact may then be written as 
\begin{equation}
\mathcal{C}(x,x_I, o_I) \lambda = 0, \ \  \mathcal{C}(x,x_I, o_I) \ge 0, \ \  \lambda(x) = \hat{n}(x) \cdot \sigma(x) \hat{n}(x)  \le 0  \quad \forall \ x\in \partial_C \Omega
\end{equation}
by introducing a Lagrange multiplier $\lambda$, with $\hat{n}$ being the outward normal to $\partial \Omega_I$.  We enforce the constraint $\mathcal{C}\ge 0$ using a slack variable $\ell$ by setting $\mathcal{C} = \ell^2$.

\paragraph{Governing equations}\label{sec:balance_laws} We derive the governing equations from the action integral
\begin{equation}
\begin{aligned}
    \widetilde{\mathcal{A}}(u, q, \varepsilon^p, \lambda) \coloneqq &\int_{t_1}^{t_2} \biggr[  \int_\Omega \left( \frac{\rho}{2} | \dot{u}|^2 \ d\Omega - \left( \frac{1}{2} \varepsilon^e \cdot \mathbb{C} \varepsilon^e + W^p(q) + \int_{0}^t \psi(\dot{q}) d\xi \right) +  b \cdot u \right) \ d\Omega \ dt  \\
    &+ \quad \int_{\partial_C \Omega} \lambda (\mathcal{C}_I(X + u, x_I) - \ell^2) \ dS \biggr] \ dt.
\end{aligned}
\end{equation}
Stationarity of this action integral over the variable set $\{u, q, \varepsilon^p, \lambda \}$ gives the governing relations:
\begin{equation} \label{eq:contact_evo}
    \begin{aligned}
    0 &= \int_\Omega \left[\rho \ddot{u} \cdot \delta u +   \mathbb{C} \varepsilon^e \cdot \nabla \delta u - b \cdot \delta u \right] \ d\Omega - \int_{\partial_C \Omega} \lambda \pdv{\mathcal{C}_I}{u} \cdot \delta u \ dS\qquad  && \forall \ \delta u \in \mathcal{K}_0, \\
    0 &\in \sigma_M - \pdv{W^p}{q} - \partial \psi(\dot{q}),  \quad  \dot{\varepsilon}^p = \dot{q} M && \text{ on } \Omega, \\
    0 &= \mathcal{C}_I(X + u, x_I) - \ell^2 && \text{ on } \partial_C \Omega, \\
    u &=u_0 && \text{ on } \partial_u \Omega, \\
    & \hspace{-1em} u|_{t = 0} = \dot{u}|_{t = 0} = 0, \ q|_{t = 0} = 0,\ \varepsilon^p|_{t = 0} = 0 && \text{ on } \Omega
    \end{aligned}
\end{equation}
at each $t \in (0,T)$ where $\mathcal{K}_0 \coloneq \{\delta u: \delta u = 0 \text{ on } \partial_u\Omega\}$. Here, $M = \sqrt{\frac{3}{2}}\frac{s}{\norm{s} }$ is plastic flow direction and $\sigma_M = \sqrt{\frac{3}{2} s \cdot s }$ is the Von Mises stress, with $s = \mathbb{C} \varepsilon^e - (1/N) \Tr(\mathbb{C} \varepsilon^e)$ being the deviatoric stress. Above, the first equation is the balance of linear momentum in weak form, the equations on the second line are the yield condition and plastic flow rule, the third is the contact constraint, and the remaining are the displacement boundary conditions and initial conditions.

\paragraph{Numerics}
We spatially discretize the domain $\Omega$ with hexahedral elements, and use first order ($Q=1$) Lagrange polynomial shape functions for the interpolation for the displacement field $u$.   The plastic quantities $q$ and $\varepsilon^p$ are spatially discretized at quadrature points.  We adopt a staggered update scheme through a predictor-corrector algorithm for contact~\cite{CAMACHO1997269,Molinari2002}. The predictor for displacements and velocities are computed explicitly through a central difference scheme assuming no contact. If penetration is detected, a correction force is computed and applied to maintain the contact condition.  Finally, the plastic quantities are updated with an implicit backwards Euler scheme. The full details are given in Appendix \ref{ap:a1}.

\subsection{Indirect inverse problem of parameter identification}
\label{sec:opt_for_viscoplastic_param}

\paragraph{Optimization Problem}

We formulate the inverse problem for finding the elasto-viscoplastic material parameters $P \coloneq \{\sigma_y,\ \varepsilon_0^p,\ n,\ \dot{\varepsilon}_0^p,\ m \}$ as an optimization problem. We are given experimental data $D^\text{exp}$, which in our case consists of the temporal evolution of the measured reaction force on the indenter and its position. We seek to find the parameters $P$ that minimize an objective 
\begin{equation} \label{eq:int_obj}
    \mathcal{O}(P, u, q, \varepsilon^p, \lambda, D^{\text{exp}}) = \int_0^T \int_\Omega o(P, u, q, \varepsilon^p, \lambda, D^{exp}) \ d\Omega \ dt,
\end{equation}
that measures the difference between the $D^\text{exp}$ and the corresponding quantities computed by a solution of the governing equations (\ref{eq:contact_evo}) with parameters $P$,
\begin{equation} \label{eq:plast_opt}
\begin{aligned}
    &\inf_{P \in \mathcal{P}} \ \mathcal{O}(P, u, q, \varepsilon^p, \lambda, D^{exp}) \\
    & \text{Subject to: } \{u, q, \varepsilon^p, \lambda \} \text{ satisfying } \eqref{eq:contact_evo} \text{ with } P.
\end{aligned} 
\end{equation}
We take the objective to be of a general integral form for now and specialize to a specific form in the next section. 

\paragraph{Adjoint Method}
We seek to solve the optimization problem using a gradient-based approach, and use the adjoint method to find the sensitivity of the objective with respect to the parameters.
We introduce the set of adjoint variables $\{v, \gamma, \zeta, \tau \}$ associated, respectively, with $\{u, q, \varepsilon^p, \lambda \}$, respectively.  The derivation is given in Appendix~\ref{ap:a2}. We note that as in~\cite{Akerson2023,AkersonLiu2023}, we use the necessary Kuhn-Tucker conditions in deriving the adjoint relations. We obtain the sensitivity,
\begin{equation} \label{eq:sensitivities}
    \dv{\mathcal{O}}{P} = \int_{0}^T  \int_{\Omega} \left[ \pdv{o}{P} + \pdv{\mathbb{C}}{P} \varepsilon^e \cdot \nabla v  + \gamma \dot{q} \left( \pdv{{\sigma}_M}{P} - \pdv{^2 W^p}{q \partial P} - \pdv{^2 {g}^*}{\dot{q} \partial P}\right)  \right]  \ d\Omega \ dt,
\end{equation}
where the adjoint variables $v$, $\gamma$, $\zeta$, and $\tau$  satisfy the adjoint equations,
\begin{equation} \label{eq:adj_rels}
    \begin{aligned}
    &0 =\int_{\Omega} \left[ \rho \ddot{v} \cdot \delta u +  \left(  \mathbb{C} \nabla v +  \gamma \dot{q} \pdv{\sigma_M}{\varepsilon} - \dot{q} \zeta \cdot \pdv{M}{\varepsilon} \right) \cdot \nabla \delta u  + \pdv{o}{u} \cdot \delta u \right] \ d\Omega \\
    &\qquad + \int_{\partial_C \Omega} \left( \tau \lambda \pdv{\mathcal{C}_I}{u}   - \lambda v \cdot \pdv{^2 \mathcal{C}_I}{u \partial u} \right) \cdot \delta u \ dS   && \forall \delta u  \in \mathcal{K}_0, \\
    & \dv{}{t} \left[ \gamma \left( \sigma_M - \sigma_0 - \pdv{g^*}{\dot{q}}\right) - \gamma \dot{q} \pdv{^2 g^*}{\dot{q}^2} - \zeta \cdot M \right]  = \pdv{o}{q} - \gamma \dot{q}  \pdv{^2 W^p}{q^2} && \text{ on } \Omega,\\
    & \dv{\zeta}{t} =  \pdv{o}{\varepsilon^p}- \mathbb{C} \nabla v + \gamma \dot{q} \pdv{\sigma_M}{\varepsilon^p} - \dot{q} \zeta \cdot \pdv{M}{\varepsilon^p} && \text{ on } \Omega, \\
    & 0 = \pdv{\mathcal{C}_I}{u} \cdot v - \pdv{o}{\lambda} && \text{ on } \partial_C \Omega_{\lambda \neq 0}, \\
     & v|_{t = T} = \dot{v} |_{t = T} = 0, \ \gamma |_{t = T} = 0, \quad  \mu |_{t = T} = 0 
    \end{aligned}
\end{equation}
at each $t \in (0,T)$.  Here, we may interpret $\tau \lambda $ as a Lagrange multiplier enforcing the constraint that $ \pdv{C_I}{u} \cdot v = \pdv{o}{\lambda}$ on regions of the boundary with nonzero $\lambda$. As in~\cite{AkersonLiu2023}, we neglect the variation of the contact area when deriving these adjoint relations. {This assumption holds except for special cases in the continuous setting. However, after the discretization of the forthcoming section, these issues are alleviated. This is discussed further in Appendix~\ref{ap:a2}.} Note that these equations specify the final conditions.  Thus, they are solved backward in time starting from the final conditions.

\paragraph{Numerics}

The spatial discretization for each of the adjoint variables are equivalent to that of their associated forward variables. We use analogous temporal discretizations to the forward problem, accommodating the backwards in time nature of the adjoint problem. Similar to the displacement field $u$, the adjoint displacements $v$ are updated with an explicit central difference method. Then, the adjoint Lagrange multiplier that satisfies the adjoint constraint is solved through a predictor-corrector scheme. Finally, the adjoint plastic variables are solved implicitly through a forward Euler scheme. The full details of this can be found in Appendix~\ref{ap:a3}.

Following the solution of the forward problem~\eqref{eq:contact_evo} and the adjoint problem~\eqref{eq:adj_rels}, the objective and  sensitivities are computed from~\eqref{eq:sensitivities}. Both are approximated with a simple Riemann sum in time and integrated with Gauss-quadrature in space. The parameters are updated using the gradient-based Method of Moving Asymptotes~\cite{svanberg1987}. This is implemented in the open-source C++ deal.ii finite element library~\cite{Arndt2021}.

\section{Demonstration using synthetic data}
\label{sec:ex}

We demonstrate our method with synthetic data in this section, and experimental data in the subsequent section.  Synthetic data provides a notion of ground truth, and thus enables the verification of the method.  Experimental data has noise, and thus demonstrates robustness.

\subsection{Setup}

The schematic of a typical test is shown in Figure~\ref{fig:indent_diagram}. {A rigid indenter with centroid $x_I(t)$ is brought into contact with a circular cylindrical domain $\Omega$ of radius $R_C=1.5$~mm and height $h=1$~mm}.  We assume that $x_I(t)$ is given, and we measure the total vertical reaction force $F_\text{exp}(t)$ on the indenter.  While it is possible to measure the residual indentation profile after the indenter is withdrawn, and the formulation in the previous section can account for that, we do not consider this information in this and the following section for two reasons.  First, the measurement of the profile is experimentally challenging, and second, we find good inference of the elastic-plastic behavior with only the reaction force measurement.

We conduct $r$ tests at various imposed indenter velocities, and consider the objective
\begin{equation}
\mathcal{O} = \sum_{i = 1}^r  \frac{\norm{\widetilde{F}^\text{P}_i - \widetilde{F}^\text{exp}_i}^2_{L^2(0, T_i) }}{\norm{\widetilde{F}^{exp}_i}^2_{L^2(0, T_i)}}  = \sum_{i = 1}^r  \frac{\int_0^{T_i} \left|\widetilde{F}^\text{P}_i - \widetilde{F}^\text{exp}_i \right|^2 dt}{\int_0^{T_i} \left| \widetilde{F}^\text{exp}_i \right|^2 dt},
\end{equation}
where ${F}^\text{exp}_i$ are the experimental reaction forces, and ${F}^\text{P}$ are computed as {${F}^\text{P} \coloneqq \int_{\partial_C \Omega} -\lambda 
\pdv{\mathcal{C}}{u}\cdot e_3 \ dS $} using the solutions from~\eqref{eq:contact_evo} with parameters $P$, with $e_r$ being the outward normal to the indenter.  The tilde denotes that filtering is applied to this quantities, and this is discussed in Appendix~\ref{ap:a1_5}. We emphasize that this objective is of a different form than the double integral expression of~\eqref{eq:int_obj}. However, the terms that replace the objective derivatives that appear in the adjoint relations~\eqref{eq:adj_rels} can be found by a straightforward application of the chain rule.

\subsection{Demonstration}


We consider a spherical indenter of radius $R_I=0.35$~mm and generate synthetic data $\{(x_I)_i, \widetilde{F}^{\text{exp}}_i \}_{i=1}^r$ by imposing $x_I(t)$ and solving the forward problem~\eqref{eq:contact_evo} for $\widetilde{F}^{\text{exp}}(t)$ with the plastic parameters $P^\text{gen}$ shown in Table~\ref{tab:params_full_table}, and use a shear modulus of $\mu = 46.7$~GPa and a Poisson's ratio $\nu = 0.3656$.   The computations are conducted on a mesh of 1580 elements, with higher refinement near the indentation site. We use $r=4$ tests with constant indenter velocities of $1.56$, $6.25$, $25$, and $100$ m/s, with corresponding timestep sizes of $\Delta t = 4 \times 10^{-3},\  2 \times 10^{-3}, \ 1 \times 10^{-3}, $ and $2.5 \times 10^{-4} \ \mu$s. For each of the tests, we indent the specimen to a total of depth of $\delta = 10\ \mu$m.

\begin{figure}[htb!]
    \centering
    \includegraphics[width=0.9\textwidth]{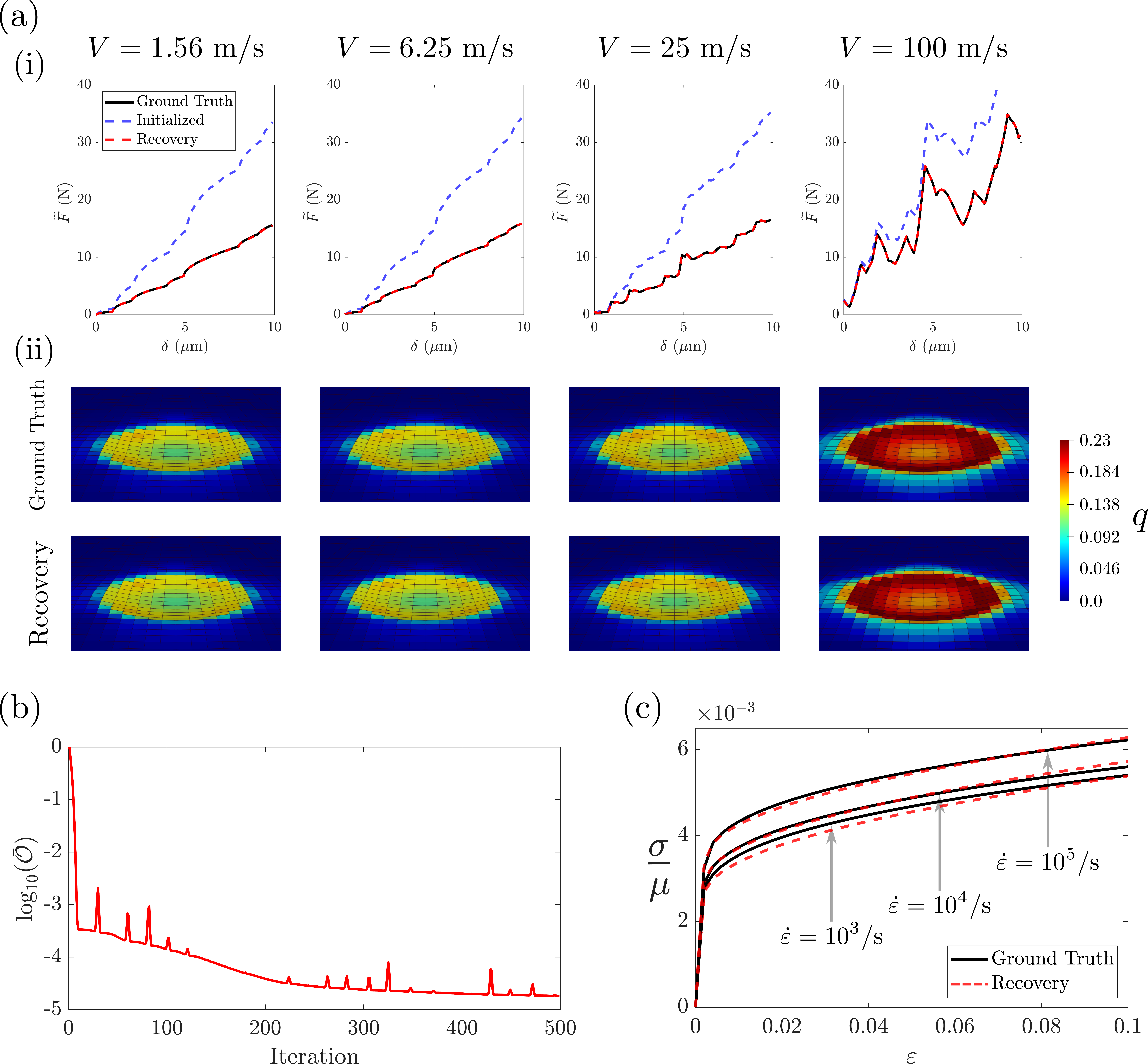}
    \caption{Demonstration with synthetic data for the the data generated with $P^{\text{gen}}_1$, with the optimization initialized from $P^{\text{init}}_1$. (a) (i) Filtered force vs indentation depth  for the synthetic, initialized, and learned parameters for the four indenter velocities tested. (a)(ii) The value of the hardening variable at the final timestep for the indentation tests at the contact site for the synthetic and learned parameters sets. For visualization purposes, these were conducted on a mesh with one level more refinement than the optimization was conducted on. (b) Normalized objective vs iteration.  (c) Comparison of synthetic and learned models in an independent test of uniaxial tension at various strain rates.    } 
    \label{fig:syn_compare}
\end{figure}
\begin{table}[htb!]
\begin{center}
\begin{tabular}{ |p{1cm} ||p{1.5cm}|p{1.8cm}|p{1.4cm}|p{1.8cm}|p{1.4cm}|p{2.2cm}|}
\hline
\multicolumn{7}{|c|}{\textbf{Synthetic and Converged Elasto-Viscoplastic Material Parameters}} \\
\hline
\ & $\sigma_y/\mu$ & $\varepsilon^p_0$ & $n$  & $\dot{\varepsilon}^p_0$  ($/s$) & $m$ & ${\mathcal{O}}$\\
\hline
$P^\text{gen}$ & 0.001935 &  0.02245&  3.23 & 5.00$\times10^5$ & 2.00 & -- \\
\hline
$P^\text{init}_1$ & 0.005000 & 0.05000 & 5.00 & 3.00$\times10^5$& 5.00 & --  \\
$P^\text{rec}_1$ & 0.001832 & 0.03878  & 2.37 & 3.70$\times10^5$& 4.50 & 3.64$\times 10^{-5}$ \newline (1.82$\times 10^{-5}$) \\
\hline
$P^\text{init}_2$ & 0.01 & 0.1 & 2.00 & 2.00$\times10^5$& 6.00 & --  \\
$P^\text{rec}_2$ & 0.001919 & 0.01894  & 3.54 & 3.37$\times10^5$& 1.54 & 5.53$\times 10^{-6}$ \newline (7.81$\times 10^{-7}$) \\
\hline
$P^\text{init}_3$ & 0.007500 & 0.07500 & 4.00 & 4.00$\times10^5$& 1.00 & --  \\
$P^\text{rec}_3$ & 0.002386 & 0.06619  & 2.56 & 3.72$\times10^5$& 5.19 & 1.45$\times 10^{-5}$ \newline (5.55$\times 10^{-6}$) \\
\hline
\end{tabular}
\end{center}
\caption{Parameters used to generate synthetic data ($P^{\text{gen}}$), initialize the algorithm ($P^{\text{init}}$), and the recovered values at the end of the iterative process ($P^{\text{rec}}$). For the recovered parameters, the objective $\mathcal{O}$ is recorded with their values normalized by the initial objective shown in parenthesis.}
\label{tab:params_full_table}
\end{table}

Given the synthetic data $\{(x_I)_i, \widetilde{F}^{\text{exp}}_i \}_{i=1}^r$, we use the method described earlier to infer the parameters from the force measurement starting from the parameters $P^\text{init}_1$ shown in Table~\ref{tab:params_full_table}.   We iterate until the relative value of the objective (normalized by the initial objective value) decreases by more than six orders of magnitude, or after 500 iterations.

The recovered parameters on convergence, $P^\text{rec}_1$, are shown in Table~\ref{tab:params_full_table}. Figure~\ref{fig:syn_compare}(a)(i) compares the total force versus indentation depth curves of the data (ground truth, black lines),  those obtained with the initialized parameters (initialized, blue dashed lines) and those obtained with the final parameters (recovery, red dashed lines), while Figure~\ref{fig:syn_compare}(a)(ii) compares the final profiles of the hardening parameters. We see excellent agreement for both.  The objective decreases with iterations as shown in Figure~\ref{fig:syn_compare} (b), falling by $4.7$ orders of magnitude over $500$ iterations.  This is more iterations than was required by both the quasistatic ($300$) and the dynamic annular compression ($150$) experiments studied in~\cite{akerson2024learning}.  Similar to~\cite{akerson2024learning}, we notice that the generating and converged parameters differ. However, the objective function is unable to distinguish between these two sets.  We further compare the model used to generate the data, and the learned model in an independent quasistatic uniaxial tensile test conducted at different strain rates, shown in Figure~\ref{fig:syn_compare}(c).  We see very good agreement, though the responses differ slightly at the lowest strain rate of $10^3$ /s. Thus, the uniaxial test is barely able to distinguish between these two sets of parameters. 

\begin{figure}
    \centering
    \includegraphics[width=0.9\textwidth]{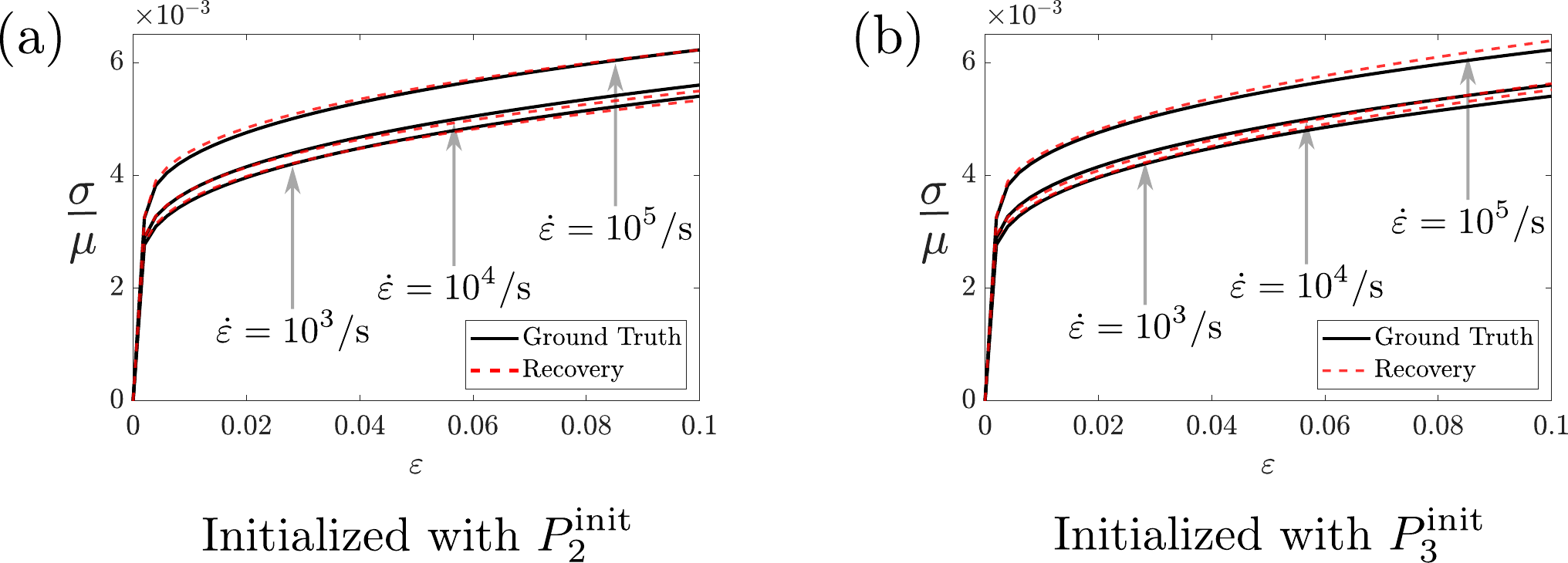}
    \caption{Response to an independent uniaxial tensile test using the synthetic parameters $P^{\text{gen}}$ and the converged parameters. (a) Response for $P^{\text{rec}}_2$. (b) Response for $P^{\text{rec}}_3$.    } 
    \label{fig:param_compare}
\end{figure}

To demonstrate the method's robustness to the choice of initial guess, we consider three different initial parameter sets $\{P^{\text{init}}_1, \ P^{\text{init}}_2, \  P^{\text{init}}_3 \}$. These, along with the corresponding recovered values, $\{P^{\text{rec}}_1, \ P^{\text{rec}}_2, \  P^{\text{rec}}_3 \}$,  are shown in Table~\ref{tab:params_full_table}. We observe that the objective decreases by over five orders of magnitude from the initial values for all of these cases. The recovered values differ considerably from each other and the set used to generate the data. However, the response to an independent uniaxial tensile test is nearly indistinguishable, as shown in Figure~\ref{fig:param_compare}. 

\begin{figure}[t]
    \centering
    \includegraphics[width=0.9\textwidth]{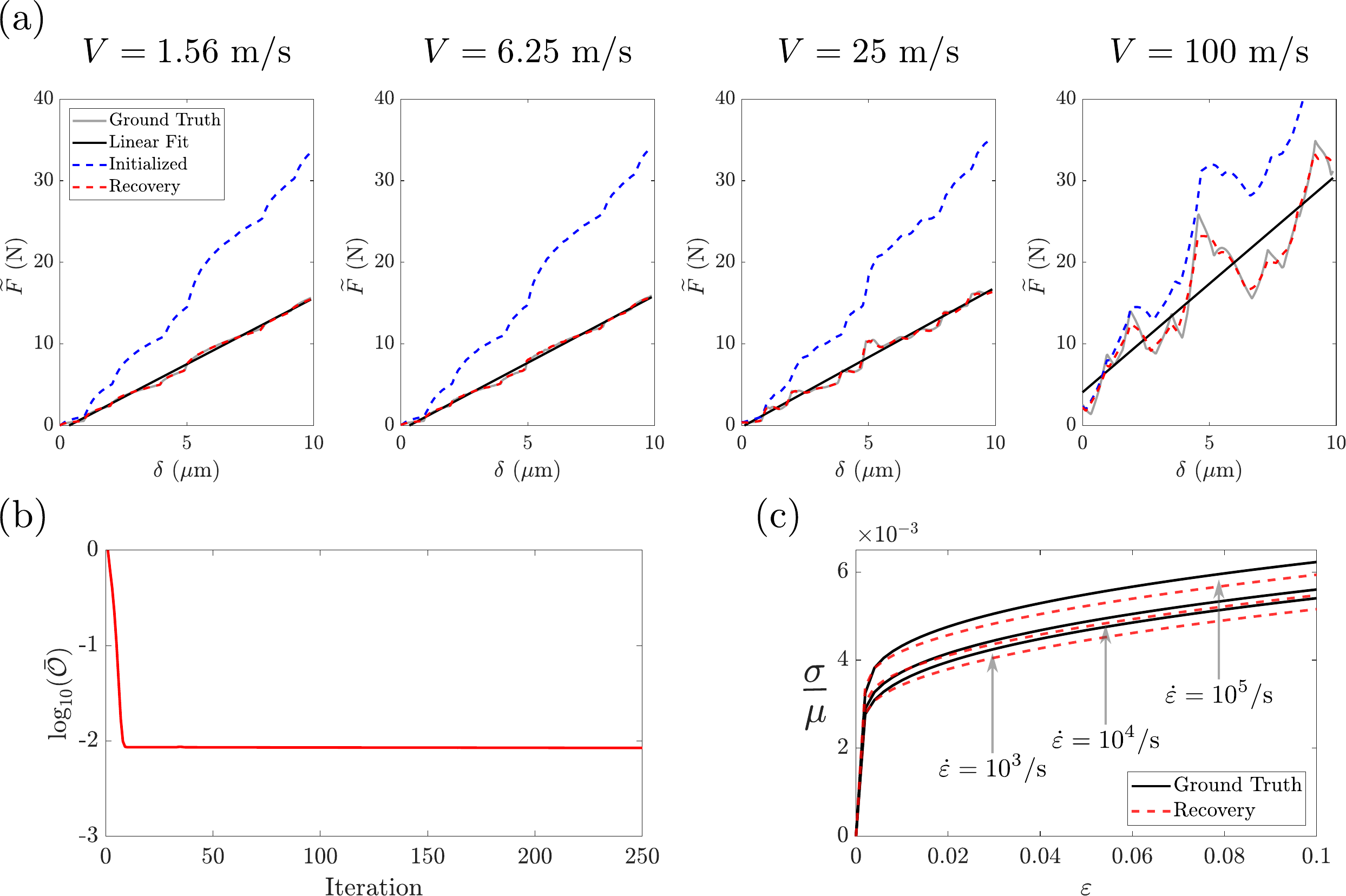}
    \caption{Demonstration on linearized synthetic data generated from $P^{\text{gen}}_\text{lin}$. (a) Filtered forces vs indentation depths for the original synthetic parameters (solid grey lines), linearization (solid black lines), initialized (blue dashed lines), and learned (dashed red lines) parameters for the four constant indenter velocities tested.(b) Normalized objective vs iteration. (c) Comparison of synthetic and learned models in an independent test of uniaxial tension at various strain rates. } 
    \label{fig:lin_compare}
\end{figure}

It is somewhat unexpected that we can recover an accurate representation of the material response using only the macroscopic contact force and indenter displacement evolutions. We hypothesize that this richness in the data arises from the often-neglected fluctuations in the dynamic response. To investigate this, we conduct a test using significantly smoothed data. Here, after generating the synthetic data, we apply a linear fit to the force-displacement response. This linearized data is fed to the optimization as $\widetilde{F}^{\text{exp}}$. The initialized parameters $P^{\text{init}}_{\text{lin}}$, and the recovered parameters $P^{\text{rec}}_{\text{lin}}$ are shown in Table~\ref{tab:params_lin}. Figure~\ref{fig:lin_compare}(a) shows the ground truth (grey solid line), linear fit (black solid line), initialized (blue dashed line), and recovered (red dashed line) force vs indentation depth curves. We observe that the recovered response does not match the oscillations in the original data, especially at the larger indentation rate. Figure~\ref{fig:lin_compare}(b) shows the objective through the iterations, and we see that it stagnates after rapidly decreasing by two orders of magnitude. Finally, Figure~\ref{fig:lin_compare}(c) shows the stress-strain responses for the uniaxial tensile test at various rates for the synthetic and recovered parameter sets. We see that the responses are close; however, it is a noticeably worse fit than the previously presented cases where we did not linearize the data. {We hypothesize that these fluctuations, that arise from the dynamic material response during the indentation event at high rates, contains significant information about the constitutive response. Including these effects in the synthetic data notably improves the fit, and allows adequate recovery from only the macroscopic indentation load and displacement response.}

\begin{table}
\begin{center}
\begin{tabular}{ |p{1cm} ||p{1.5cm}|p{1.8cm}|p{1.4cm}|p{1.8cm}|p{1.4cm}|p{2.2cm}|}
\hline
\multicolumn{7}{|c|}{\textbf{Recovery of Elasto-Viscoplastic Material Parameters from Linear Fit}} \\
\hline
\ & $\sigma_y/\mu$ & $\varepsilon^p_0$ & $n$  & $\dot{\varepsilon}^p_0$  ($/s$) & $m$ & ${\mathcal{O}}$\\
\hline
$P^\text{gen}_{\text{lin}}$ & 0.001935 &  0.02245&  3.23 & 5.00$\times10^5$ & 2.00 & -- \\
\hline
$P^\text{init}_{\text{lin}}$ & 0.005000 & 0.05000 & 5.00 & 3.00$\times10^5$& 5.00 & --  \\
$P^\text{rec}_{\text{lin}}$ & 0.001777 & 0.04573  & 2.56 & 3.72$\times10^5$& 5.91 & 1.71$\times 10^{-2}$ \newline (8.45$\times 10^{-3}$) \\
\hline
\end{tabular}
\end{center}
\caption{Recovery from the linearized data. Parameters used to generate synthetic data ($P^{\text{gen}}_{\text{lin}}$) initialize the algorithm ($P^{\text{init}}_{\text{lin}}$), and the recovered values at the end of the iterative process ($P^{\text{rec}}_{\text{lin}}$). For the recovered parameters, the objective $\mathcal{O}$ is recorded with the value normalized by the initial objective shown in parenthesis.}
\label{tab:params_lin}
\end{table}

\begin{figure}[ht]
    \centering
    \includegraphics[width=1.0\textwidth]{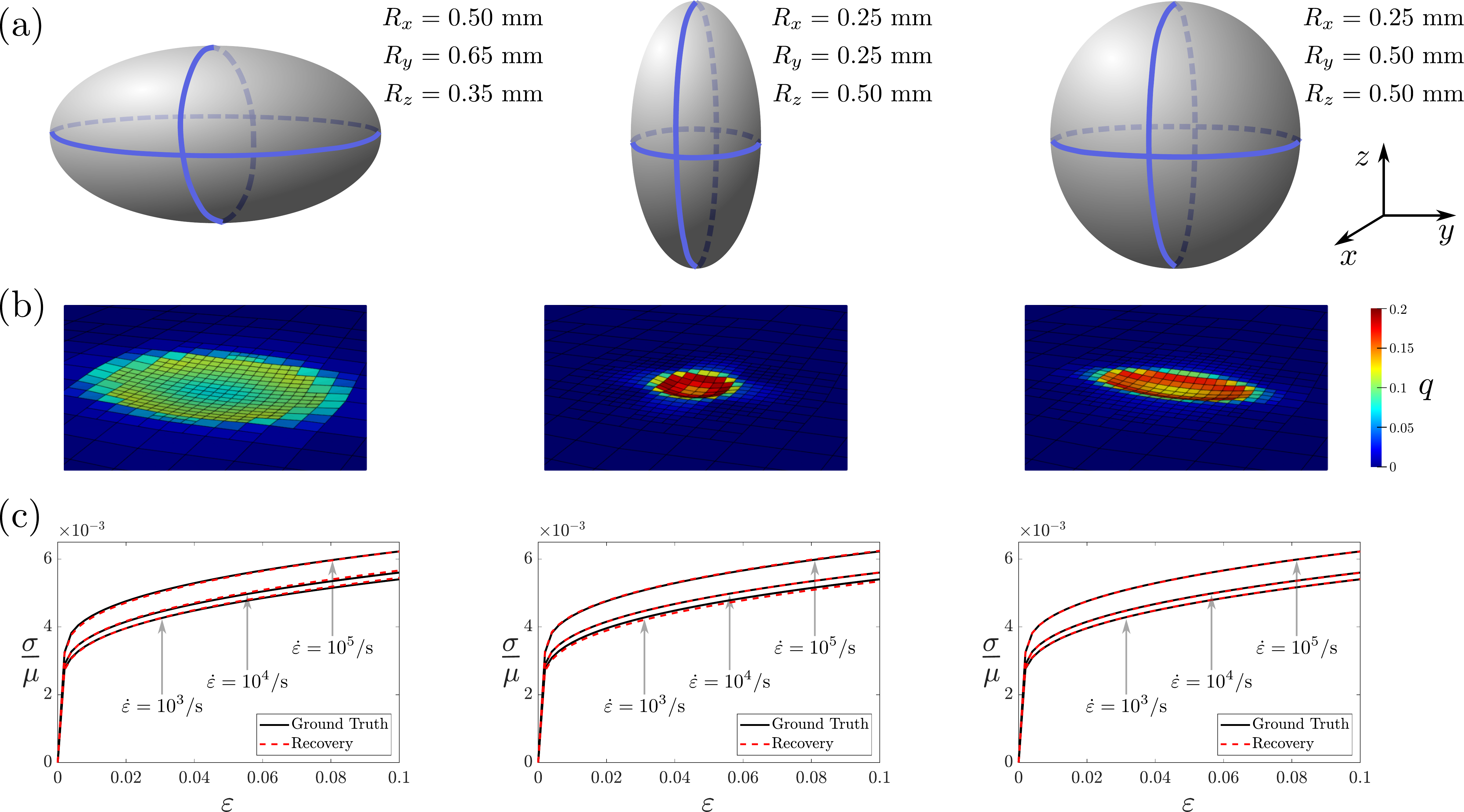}
    \caption{  Demonstration on ellipsoidal indenters with different dimensions. (a) Diagrams of the ellipsoidal morphologies and their corresponding dimensions. (b) Snapshots of the indentation profile and accumulated plasticity at the final timestep for the simulation using a 25~m/s indenter speed. (c) Stress-strain curves for independent uniaxial tension tests using the recovered parameters (red dashes), and the ground truth parameters used to generate the synthetic data (black lines). } 
    \label{fig:ellipsoid_rec}
\end{figure}

{
We demonstrate efficacy of the method for indenters of different shapes. The non-smooth features of conical or Vickers indentation can cause numerical issues in resolving the contact force direction. Thus, we consider ellipsoidal indenters of semi-axis lengths $\{R_x, R_y, R_z \}$. In this setting, the contact algorithm must be modified from the specialized spherical case, and this is detailed in Appendix~\ref{ap:a4}. For all of the ellipsoidal cases, we consider an identical domain to the previous spherical studies with the same elastic properties. However, we use a finer mesh of 3064 elements to resolve contact with ellipses having sharper features. We use $r=4$ tests with constant indenter velocities of $1.56$, $6.25$, $25$, and $100$ m/s, with corresponding timestep sizes of $\Delta t = 2.29 \times 10^{-3},\  1.33 \times 10^{-3}, \ 6.67 \times 10^{-4}, $ and $1.67 \times 10^{-4} \ \mu$s.

We consider three cases of ellipsoidal indenters with varied dimensions, as shown in Figure~\ref{fig:ellipsoid_rec}(a). In each of these, the data is generated synthetically from initial simulations using the same parameter set $P^{\text{gen}}_{\text{ellipsoid}}$. Snapshots of the resulting indented surfaces are shown in Figure~\ref{fig:ellipsoid_rec}(b). The optimization for all of the cases are initialized with identical $P^{\text{init}}_{\text{ellipsoid}}$, and the recovered parameters after 500 optimization iterations, $P^{\text{rec}}_{\text{ellipsoid, i}}$, are shown in Table~\ref{tab:params_ellipse}. We observe that the objective decreases by over 5 orders of magnitude for all of the cases. While the recovered parameters differ slightly, the recovered stress strain curves for independent uniaxial tension tests in Figure~\ref{fig:ellipsoid_rec}(c) remain nearly identical. As we have considered ellipsoidal indenters of highly varied morphologies, we conclude that the method is robust to the choice of indenter shape. 
}

\begin{table}
\begin{center}

\begin{tabular}{ |p{1.5cm} ||p{3.1cm} | p{1.5cm} |p{1.4cm}|p{0.8cm}|p{1.6cm}|p{0.8cm}|p{2cm}|}
\hline
\multicolumn{8}{|c|}{\textbf{Recovery of Elasto-Viscoplastic Material Parameters for Ellipsoidal Indenters}} \\
\hline
\ & $\{R_x,R_y,R_z \} \ (\text{mm})$ & $\sigma_y/\mu$ & $\varepsilon^p_0$ & $n$  & $\dot{\varepsilon}^p_0$  ($/s$) & $m$ & ${\mathcal{O}}$\\
\hline
$P^\text{gen}_{\text{ellipsoid}}$ & -- & 0.001935 &  0.02245&  3.23 & 5.00$\times10^5$ & 2.00 & -- \\
$P^\text{init}_{\text{ellipsoid}}$ & -- & 0.005 &  0.05 &  5.0 & 3.00$\times10^5$ & 5.00 & -- \\
\hline
$P^\text{rec}_{\text{ellipsoid,1}}$ & $\{0.5,\ 0.65,\ 0.35 \}$ & 0.001859 & 0.01864 & 3.19 & 6.21$\times10^5$& 2.41 & 2.29$\times 10^{-6}$ \newline (1.10$\times 10^{-6}$)   \\
$P^\text{rec}_{\text{ellipsoid,2}}$ & $\{0.25,\ 0.25,\ 0.5 \}$ & 0.001838 & 0.02119  & 3.13 & 4.36$\times10^5$& 2.59 & 4.43$\times 10^{-6}$ \newline (1.81$\times 10^{-6}$) \\
$P^\text{rec}_{\text{ellipsoid,3}}$ & $\{0.25,\ 0.5,\ 0.5 \}$ & 0.001992 & 0.02624  & 3.17 & 5.08$\times10^5$& 1.89 & 1.87$\times 10^{-7}$ \newline (7.91$\times 10^{-8}$) \\
\hline
\end{tabular}
\end{center}
\caption{Recovery from ellipsoidal indenters of three different dimensions. For all of the cases, we generate synthetic data with $P^{\text{gen}}_{\text{ellipsoid}}$ and initialize the algorithm from $P^{\text{init}}_{\text{ellipsoid}}$. Recovered parameters at the end of the iterative process ($P^{\text{rec}}_{\text{ellipsoid}}$) and the corresponding ellipsoid dimensions are shown. For the recovered parameters, the objective $\mathcal{O}$ is recorded with the value normalized by the initial objective shown in parenthesis.}
\label{tab:params_ellipse}
\end{table}

We conclude that our method accurately recovers the viscoplastic material behavior from indentation tests on synthetically generated data. We emphasize that the method is robust to the initial choice of parameters {and indenter shape}, with the recovered and synthetic parameter sets having nearly identical stress-strain responses in independent uniaxial tensile tests. We perform this recovery using only the macroscopic indentation force vs displacement data, and conclude that the often-neglected fluctuations in this data contains sufficient information to recover the material response.

\section{Demonstration on Experimental Data}\label{sec:expt_inv}
\subsection{Experimental setup}

We  demonstrate our method using experimental data collected from dynamic indentation tests performed on a Kolsky or split-Hopkinson pressure bar~\cite{casem2023kolsky}.   The experimental setup  consists of a striker, an input bar, the specimen, and an indenter bar that serves as the output bar.  A hemispherical indenter is milled directly onto the tungsten carbide output bar.  The input bar, the specimen and the indenter bar are placed in (stress free) contact with each other, and the striker is driven into the input bar.  A stress pulse travels through the input bar and reaches the specimen, driving it into the indenter.   A combination of interferometers and strain-gages measure the initial, reflected and transmitted pulse, and these are used to compute the indentation velocity and reaction force.  In our tests, the indentation velocity is O(1 m/s), the peak indentation force is O(100 N), and total penetration depth is O(10\ $\mu$m). Full details of the setup can be found in~\cite{casem2023kolsky}

\subsection{Rolled homogeneous armor steel}

\begin{figure}[t]
\centering
    \includegraphics[width=\textwidth]{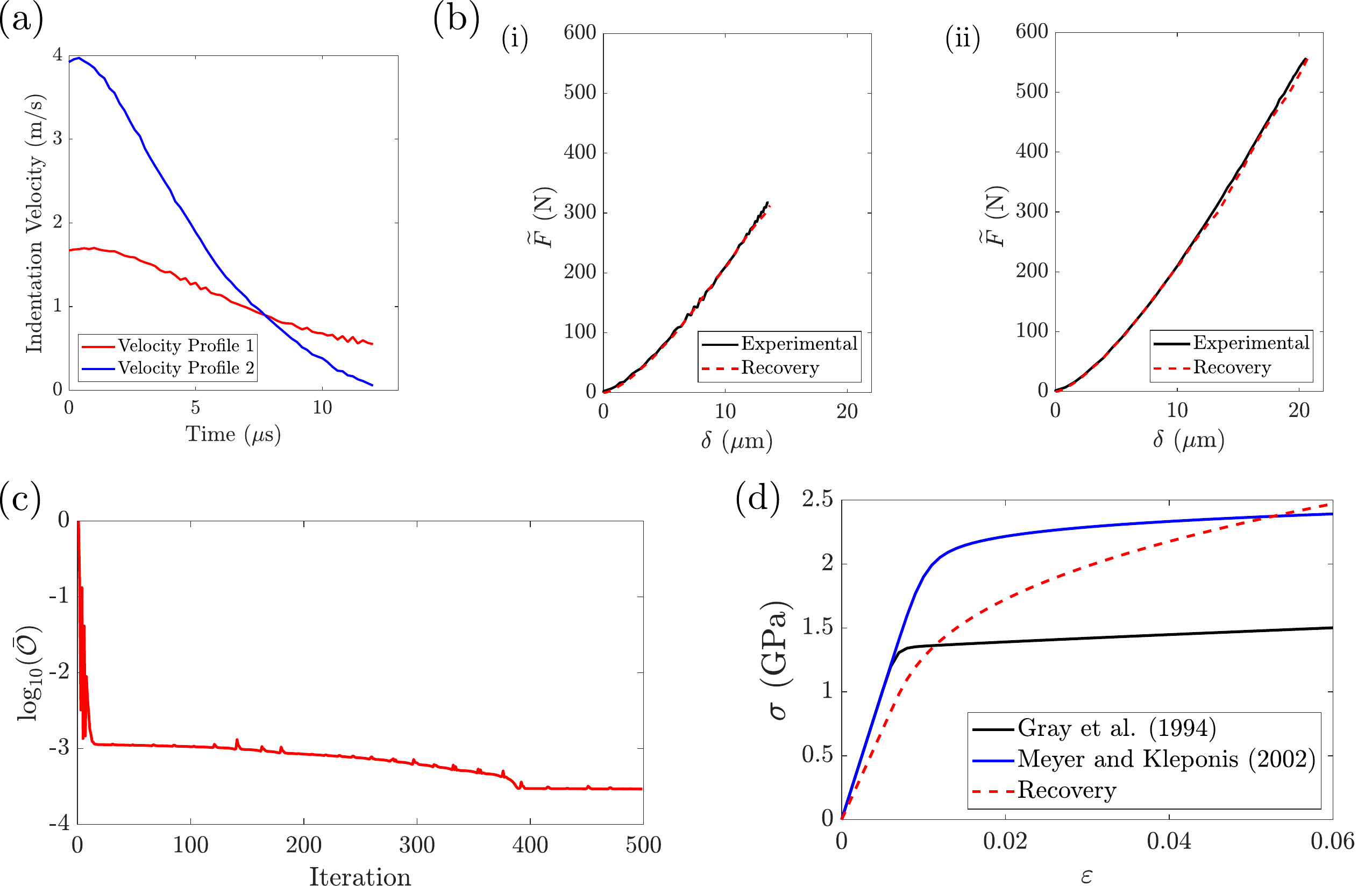}
\caption{Results for experiments with RHA steel.  (a ) Indentation velocity vs time for the two tests conducted. (b) Force versus indentation depth for (i) velocity profile 1 and (ii) velocity profile 2. (c) Normalized objective versus iteration.  (d) Comparison of the recovered model with two literature models in independent uniaxial tension tests conducted at room temperature at a strain rate of $5 \times 10^3$/s.}
\label{fig:exp_RHA}
\end{figure}

The first set of experiments are conducted on a rectangular RHA steel specimen of dimensions $10\ \text{mm} \times 10\ \text{mm} \times 4\ \text{mm}$ with a tungsten carbide indenter of radius $1530 \ \mu \text{m}$. The indentation is directed parallel to the direction of the shortest dimension ($4\ \text{mm}$). Two tests are performed with the velocity histories shown in Figure~\ref{fig:exp_RHA}(a).  The corresponding experimental measured forces are shown in Figures~\ref{fig:exp_RHA}(b)(i) and~\ref{fig:exp_RHA}(b)(ii) as black solid lines.  

We apply our method, and the recovered parameters after 500 iterations are shown in Table~\ref{tab:RHA_params_rec}. Because the experiments span limited rates, we only fit the elastic modulus, the yield stress, and the strain hardening constants while neglecting the rate parameters. Additionally, we assume a Poisson's ratio of $\nu = 0.27$ for the RHA steel specimen. The (normalized) objective decreases rapidly with iteration as shown in Figure~\ref{fig:exp_RHA}(c). We observe that the (normalized) objective does not decrease by as much as the synthetic data, likely due to the noise in the experimental data.  The contact force vs displacement curves for the recovered parameters are shown in Figures~\ref{fig:exp_RHA}(b)(i) and~\ref{fig:exp_RHA}(b)(ii) as dashed red lines.  We see excellent agreement in the modeled and experimental curves for both tests.

\begin{table}[t]
\begin{center}
\begin{tabular}{ |p{1cm} ||p{1.5cm}|p{1.8cm}|p{1.8cm}|p{1.8cm}|p{1.8cm}|p{2.2cm}|}
\hline
\multicolumn{6}{|c|}{\textbf{Recovery of Elastic-Plastic Parameters from Experimental Data}} \\
\hline
\ & $E$ (GPa)  & $\sigma_y$ (GPa) & $\varepsilon^p_0$ & $n$  & ${\mathcal{O}}$\\
\hline
\multicolumn{6}{|c|}{\textbf{RHA Steel}} \\
\hline
$P^\text{init}$ & 250.0 &  1.97 & 1.0 & 10 & --\\
\hline
$P^\text{rec}$ & 139.9  & 0.88 & $9.2 \times 10^{-3}$ & 2.68 & 1.85$\times10^{-4}$ \newline (2.92$\times 10^{-4}$)  \\
\hline
\multicolumn{6}{|c|}{\textbf{Aluminum Alloy Al~6061-T6}} \\
\hline
$P^\text{init}$ & 85.0 &  2.23 & 1.0 & 5.0 & --\\
\hline
$P^\text{rec}$ & 80.1  & 0.42 & 1.07 & 2.58 & 2.94$\times10^{-4}$ \newline (7.04$\times 10^{-4}$)  \\
\hline
\end{tabular}
\end{center}
\caption{Parameters used to initialize the algorithm ($P^{\text{init}}$), and the recovered values at the end of the iterative process ($P^{\text{rec}}$) for the experimental data on RHA steel and aluminum alloy Al~6061-T6. For the recovered parameters, the objective $\mathcal{O}$ is recorded with their values normalized by the initial objective shown in parenthesis.}
\label{tab:RHA_params_rec}
\end{table}

We compare our results with literature-reported characterization of RHA steel~\cite{gray1994,meyer2001analysis}. Here, high strain-rate compression data is used to fit the Johnson-Cook plasticity model with flow stress
\begin{equation}
    Y = \left[ A + B (\varepsilon^p)^{n_{jc}} \right] \left[ 1 + C \log(\dot{\varepsilon}^p) \right] \left(1 - \bar{\theta}^{m_{jc}} \right),
\end{equation}
where $\bar{\theta}$ is the relative temperature. The model is characterized by the set of parameters $\{A, B, n_{jc}, C, m_{jc} \}$. As these tend to vary drastically throughout the literature, we show two sets of reported values that were fit to experimental data in Table~\ref{tab:JC_params}. We compare the response of our recovered model to that of these two Johnson-Cook models in an independent quasistatic uniaxial tension test. These are simulated at room temperature at a strain-rate of $5 \times 10^3$/s, which is approximately the maximum strain rate observed in the indentation tests. Figure~\ref{fig:exp_RHA}(d) shows the stress-strain responses for these tests.  The response of our model in the indentation test reach a maximum plastic strain of approximately $6\%$. Thus, our model is only fit on data for strains lower than this value. Within this range, we observe that the response of our model falls within the bounds of the two Johnson-Cook models from the literature.

\begin{table}[t]
\begin{center}
\begin{tabular}{ |p{5cm} ||p{1.5cm}|p{1.5cm}|p{1.5cm}|p{1.0cm}|p{1.3cm}|p{1.2cm}|}
\hline
\multicolumn{7}{|c|}{\textbf{Literature: High Strain-Rate Johnson-Cook Characterization}} \\
\hline
\ & $E$ (GPa)  & A (GPa) & B (GPa) & $n_{jc}$  & C & $m_{jc}$ \\
\hline
RHA Steel: Gray et al.  & 200.0 &  0.9 & 1.305 & 0.9  & 0.0575 & 1.075\\
\hline
RHA Steel: Meyer \& Kleponis & 200.0  & 0.78 & 0.78 & 0.106 & 0.0891 & 1.00 \\
\hline
Al 6061-T6: Clayton et al. & 71.0  & 0.3 & 0.2 & 0.3 & 0.05 & 1.00 \\
\hline
\end{tabular}
\end{center}
\caption{Parameter fits to high strain-rate compression data on RHA steel and aluminum alloy Al~6061-T6 using the Johnson-Cook plasticity model taken from Gray et al.~\cite{gray1994}, Meyer and Kleponis~\cite{meyer2001analysis}, and Clayton et al.~\cite{clayton2023simulation}.}
\label{tab:JC_params}
\end{table}

We comment on the discrepancy between our recovered elastic modulus of $E^{\text{rec}} = 139.9$~GPa and the reported value of $E^{RHA} = 200$~GPa for RHA steel. We suspect there are two possible reasons for this. In the indentation tests performed here, where the loading is monotonically increasing, the elastic regime is small. Thus, most of the data from the experiment is well within the plasticity regime of the material and we  expect a less accurate fit for the elastic modulus.

Another possible source to the discrepancy is the rigid contact assumption. Our model considers a perfectly rigid indenter with no deformation. However, the tungsten carbide indenter used in the experiment was measured to have an elastic modulus of $E^{WC} = 575$~GPa. Thus, we expect it to experience measurable deformation upon indentation with RHA steel. Computing the analogous Hertz contact effective modulus for rigid contact in this scenario~\cite{HertzHeinrich1882UdBf,Johnson_1985}, 
\begin{equation}
    E^{\text{eff}} = \left(1 - \nu_{RHA}^2 \right)\left( \frac{1 - \nu_{RHA}^2}{E^{RHA}} + \frac{1 - \nu_{WC}^2}{E^{WC}}\right)^{-1},
\end{equation}
using Poisson's ratios $\nu_{RHA} = 0.27$ and $\nu_{WC} = 0.31$ and $E^{RHA} = 200$~GPa gives $E^{\text{eff}} = 149$~GPa. This is within $7\%$ of our recovered value. Thus, the difference between the recovered and widely accepted modulus may also arise from the rigid contact assumption.

{ We considered alternative approaches for the recovery, and we briefly discuss them. The first alternative is modifying the objective to weigh the low-stress regimes more heavily. However, this is not expected to improve the recovery of elastic properties, as we already observe an excellent match to the macroscopic force-displacement curves even in the low stress regimes. Additionally, the signal-to-noise ratio of the experimental data is lower for the initial loading. Thus, the data in the low-stress regimes data has a lower accuracy. The second alternative approach would be to assume the elastic modulus, and only recover the plastic parameters. However, this would not correct the interface compliance issue that arises from the rigid indentation assumption of the model. Thus, the authors chose to recover both the elastic and plastic properties with a standard objective function. This also demonstrates that the general method may be applied to unexplored material systems, where the elastic properties are not well-characterized in the regimes of interest.}

We conclude that our method effectively characterizes the elastic-plastic behavior of RHA steel using only the macroscopic contact force and indentation depth measurements.


\subsection{Aluminum alloy Al 6061-T6}
\begin{figure}[htb!]
\centering
    \includegraphics[width=\textwidth]{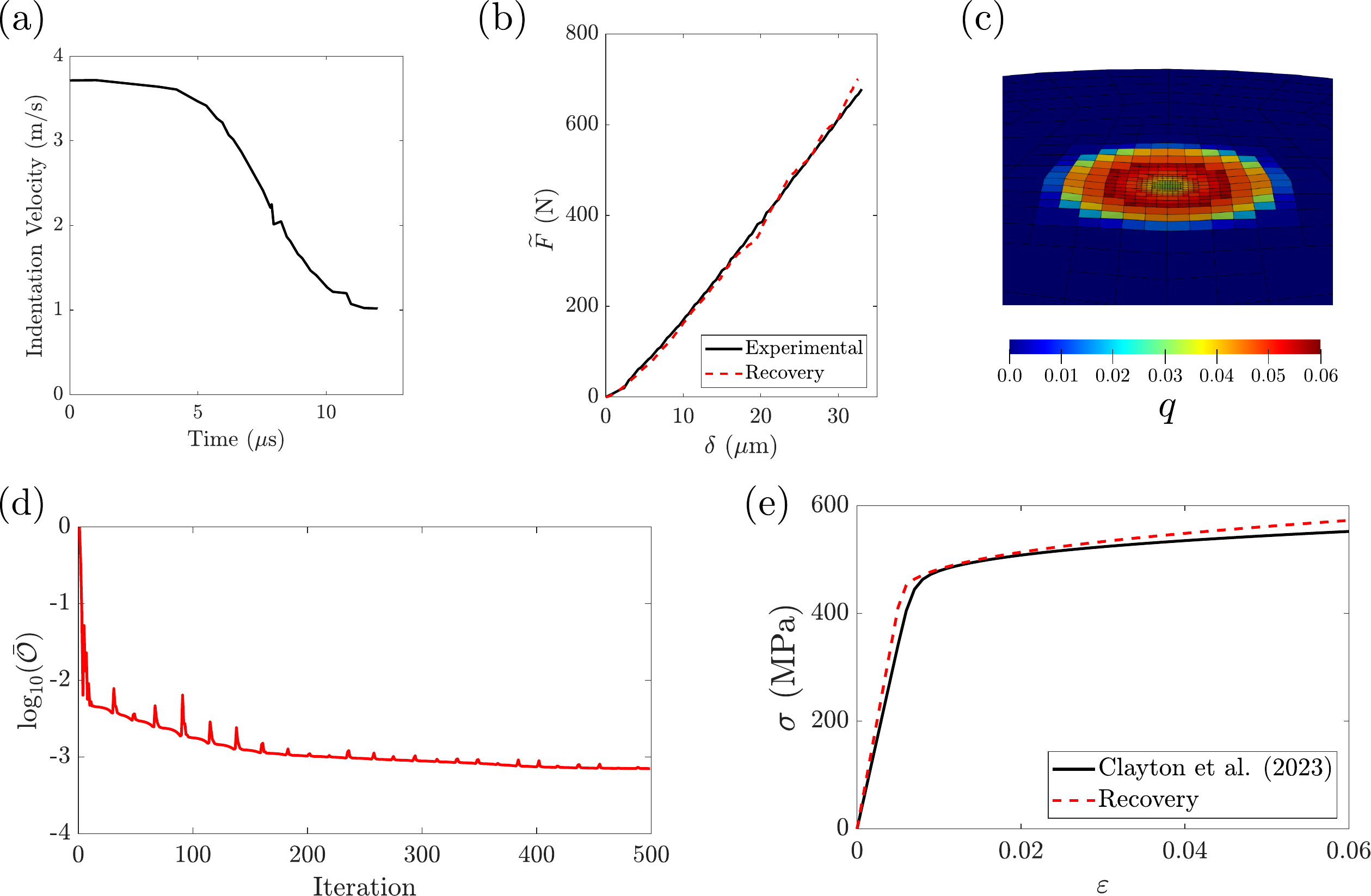}
\caption{Characterization results from experiments on aluminium alloy Al~6061-T6. (a) Indentation velocity vs time for the conducted test. (b) Measured force versus indentation depth. (c) Snapshot of the deformation field for the recovered parameters at the final timestep. (d) Objective versus iteration throughout the optimization process. (e) Comparison of the recovered model with a literature model in an independent uniaxial tension tests conducted at room temperature at a strain rate of $5 \times 10^3$/s.}
\label{fig:exp_Al}
\end{figure}

The second demonstration is performed on the indentation experiment of aluminum alloy Al 6061-T6 reported on in~\cite{clayton2023simulation}. A cylindrical specimen of radius $2.39\ \text{mm}$ and length $3.58\ \text{mm}$ is indented using a tungsten carbide indenter with a radius of $3.175\ \text{mm}$. We consider only a single velocity profile shown in Figure~\ref{fig:exp_Al}(a), resulting in a force vs indentation depth curve shown in Figure~\ref{fig:exp_Al}(b) (solid black line). 

We recover the parameters shown in Table~\ref{tab:RHA_params_rec} assuming a poissons ratio of $\nu = 0.33$. A snapshot of the deformation field for the recovered parameters at the final timestep is shown in Figure~\ref{fig:exp_Al}(c). Convergence is reached after 500 iterations, with the objective shown in Figure~\ref{fig:exp_Al}(d).  We compare the response of our recovered model with a previously calibrated Johnson-Cook model on high strain-rate data with parameters shown in Table~\ref{tab:JC_params}. Figure~\ref{fig:exp_Al}(e) shows the response of both models in an independent uniaxial tensile test conducted quasistatically at a strain rate of $5\times 10^3$/s. We see excellent agreement. { We further compare the simulated response of the recovered model to that of the literature Johnson Cook model in the indentation test, and this is detailed in Appendix~\ref{ap:a5}. The recovered model is calibrated to match this experimental curve, and thus it is unsurprisingly a much closer match to the data than the Johnson Cook model.}

We again remark on the discrepancy between the recovered elastic modulus ($80.1$\ GPa) and the widely accepted modulus ($\sim70$\ GPa). We suspect that this arises from low representation of the purely elastic response in the indentation data. 



\section{Conclusions}
\label{sec:conclusion}

Inverting constitutive relations from experimental measurements is necessary to close the continuum balance laws and perform accurate physics simulations. In part I of this work, we formulate this as a PDE constrained optimization problem where we look to find the material parameters that minimize the difference between the experimental measurements and the simulated response. This allows for parameter recovery over complex, {heterogeneous} stress-strain trajectories at the expense of solving a difficult inverse problem. In part I, we develop the formulation and solution strategy through a gradient-based optimization approach using the adjoint method to compute the sensitivities. 

In this current part II, we extend the formulation to account for dynamic rigid contact. We account for the unilateral contact constraint in the forward problem through a Lagrange multiplier and slack variable. We employ an efficient staggered numerical algorithm to compute the dynamical evolution while strictly enforcing the contact condition. The adjoint relations are derived in this setting, and we solve these with a numerical technique similar to that used for the forward problem.

We demonstrate the method on synthetic data, recovering the behavior of a J2 viscoplastic material that features power-law strain and rate hardening. The recovered parameters accurately capture the contact force vs indentation depth curves. To further validate the recovery, we conduct independent quasistatic uniaxial tensile simulations, and compare the response with that of the synthetic parameter set, where we observe excellent agreement. We additionally demonstrate that our method is robust to initial choice of parameters {and varied indenter shapes}. With only the macroscopic load vs indentation depth evolution {from a few tests}, we can accurately recover material behavior from synthetic data. {The time-series of the macroscopic data, along with the fluctuations in these curves, encodes sufficient richness for accurate characterization.}

We apply the inversion methodology to experimental indentation data. Notably, we apply this approach to both RHA-steel and Al 6061-T6 specimens, successfully recovering elastic modulus, yield, and strain hardening parameters. We verify the method by comparing the response of the recovered parameters to that of previously recorded values from the literature in an independent uniaxial tensile test. We see excellent for both materials. 

{We summarize direct extensions of this work in the context of characterization from indentation tests. The presented method is general and is readily extended to include other phenomena. This is relevant to HCP materials that are both highly anisotropic and have twinning with tension-compression asymmetry.  The general method could also be adapted to brittle materials such as ceramics. However, the approach would need to be heavily extended to account for fracture. This would be involve enriching the forward model with a fracture model (e.g. phase-field fracture~\cite{Bourdin2007}), and propagating this through the adjoint derivation. Here, special attention must be put towards the associate numerical scheme to ensure tractable run times of the optimization procedure. Finally, there are extensions that would greatly improve the robustness of the characterization. In this work we only conducted the recovery from indentation tests at varying rates. However, the optimization-based approach could be easily extended to recover a single parameter set from multiple experimental set-ups (e.g. dynamic compression, pressure-shear, uniaxial tests).}

In forthcoming part III of this work, we extend our methodology to encompass a generalized constitutive law by integrating a history-dependent neural network as the material model~\cite{ren2020benchmarking, deng2021neural, xiao2021deep}. Recently, recurrent neural operators have been shown to effectively capture the behavior of plastic materials through an arbitrarily defined set of internal variables~\cite{liu2023learning, bhattacharya2023learning, kovachki2023neural}. As the {adjoint-based characterization approach} scales nearly independently with respect to the number of parameters~{\cite{akerson2024learning}}, the computation time for this inversion remains largely unaffected by the large parameter set of the recurrent neural operator.

\section*{Acknowledgement}
This work was largely conducted with AA and AR were at the California Institute of Technology.  We are delighted to acknowledge many useful discussions with Professors Ravi Ravichandran and Andrew Stuart.  We are grateful to Professor Curt Bronkhorst of the University of Wisconsin who shared the RHA steel from which the experimental specimens were extracted.  We are grateful for the financial support of the Army Research Laboratory (W911NF22-2-0120) and the Army Research Office (W911NF-22-1-0269).


\appendix

\section{Numerical method for the governing equations} \label{ap:a1}

We detail the finite element discretization and numerical scheme we use to solve the forward problem. We consider a quadrilateral mesh with standard $Q = 1$ Lagrange polynomial shape functions for the displacement interpolation,
\begin{equation}\label{eq:dis_u}
    u = \sum_{i = 1}^{n_u} u_i N_i(x),
\end{equation}
where $N_i : \Omega \to \mathbb{R}^n$ are the standard vector-valued shape functions with compact support. The fields $q$ and $\varepsilon^p$ are discretized at quadrature points as
\begin{equation} \label{eq:dis_plas}
    q(x_g) = q_g, \qquad \varepsilon^p(x_g) = \varepsilon^p_g,
\end{equation}
for some Gauss point $x_g$, $g = 1,\dots,n_g$. We solve the evolution equations 
\begin{equation} 
    \begin{aligned}
    0 &= \int_\Omega \left[\rho \ddot{u} \cdot \delta u +   \mathbb{C} \varepsilon^e \cdot \nabla \delta u - b \cdot \delta u \right] \ d\Omega - \int_{\partial_C \Omega} \lambda \pdv{\mathcal{C}_I}{u} \cdot \delta u \ dS\qquad  && \forall \delta u \in \mathcal{K}_0, \\
    0 &\in \sigma_M - \pdv{W^p}{q} - \partial \psi(\dot{q}) \quad && \text{ on } \Omega, \\
    \dot{\varepsilon}^p &= \dot{q} M  && \text{ on } \Omega, \\
    0 &= \mathcal{C}_I(X + u, q_I) - \ell^2 && \text{ on } \partial \Omega, \\
    & \hspace{-1em} u|_{t = 0} = \dot{u}|_{t = 0} = 0, \ q|_{t = 0} = 0,\ \varepsilon^p|_{t = 0} = 0. && 
    \end{aligned}
\end{equation}
The first equation is the dynamic evolution of the displacement field $u$. The second and third equations represent the kinetic relations for the plastic variables $q$ and $\varepsilon^p$, with the fourth line being the unilateral contact constraint. 

To solve this system numerically, we adopt a staggered solution strategy.  We use an explicit central difference scheme to update a prediction for the displacement field, assuming that contact is not present. Then, we conduct a contact detection for the nodes on the contacting surface. If we observe interpenetration, we compute a contact force and apply it to the displacement field. We then update the plasticity fields $q$ and $\varepsilon^p$ implicitly with a backwards Euler update. For the $n$ to $n+1$ time-step the predictor displacements are updated as
\begin{equation}
\begin{aligned}    
    F_i^n &= \int_{\Omega} \left[ -\mathbb{C}\varepsilon^e(u^n)\cdot \nabla N_i  + b \cdot N_i \right] \, d\Omega, \\
    \ddot{u}_i^{n, \text{pre}} &= \frac{1}{m_i} F_i^n(u^n, \varepsilon^{p,n}, q^n, t^n), \\
    \dot{u}_i^{n + 1/2, \text{pre}} &= \dot{u}_i^{n - 1/2} + \Delta t^n \, \ddot{u}_i^{n, \text{pre}}, \\
    u_i^{n + 1, \text{pre}} &= u_i^n + \Delta t^{n + 1/2} \, \dot{u}_i^{n + 1/2, \text{pre}}.
\end{aligned}
\end{equation}
where the lumped mass vector is
\begin{equation}
m_i = \sum_{j = 1}^{n_u} \int_{\Omega} \rho(x) N_i \cdot N_j \, d\Omega. 
\end{equation}
Here, $F^n_i$ is the term that contains the internal and body forces. Then, we conduct a search to see if interpenetration is detected. That is, we check if
\begin{equation}
    \mathcal{C}_I(X^i + u^{n+1, \text{pre}}(X^i), \ x_I(t)) < 0 \text{ for } X^i \in \partial_C \Omega.   
\end{equation}
If the contact condition is violated at node $i$, we compute the interpenetration distance $\delta_i$ and contact force
\begin{equation}
\begin{aligned}
    0 &=  \mathcal{C}_I(X_i + u_i^{n+1, \text{pre}}(X^i) + \delta_i \ e_r,\  x_I(t)), \\
    F^c_i &= \frac{\delta_i m_i}{\Delta t^n \Delta t^{n + 1/2}} \ e_r ,
\end{aligned}
\end{equation}
where $e_r$ is the outward unit normal of the indenter. Then, the contact forces are applied to correct the velocity, and displacement fields
\begin{equation}
\begin{aligned}
    \dot{u}_i^{n + 1/2} &= \dot{u}_i^{n + 1/2, \text{pre}} + \Delta t^n \frac{F^c_i}{m_i},\\
    {u}_i^{n+1} &= {u}_i^{n+1, \text{pre}} + \Delta t^{n + 1/2} \Delta t^n \frac{F^c_i}{m_i}.
    \end{aligned}
\end{equation}
This completes the update to the displacement and velocity fields for the timestep. We then update the plasticity variables through an implicit backwards Euler discretization. For this, we employ a predictor-corrector scheme~\cite{Ortiz1999} to solve point-wise at each quadrature point,
\begin{equation}
\begin{aligned}
& 0 \in {\sigma}_M (\varepsilon^{n + 1} |_{x_g}, \varepsilon^{p, n + 1}_g ) - \pdv{W^p}{q} \ (q^{n+1}_g ) - \partial g^* \left( \frac{q^{n + 1}_g - q^n_g}{\Delta t^n} \right), \\
& \varepsilon^{p, n+1}_g = \varepsilon^{p, n}_g + \Delta q M (\varepsilon^{n + 1}_g, \varepsilon^{p, n + 1}_g).
\end{aligned}
\end{equation} 

\section{Contact force filtering} \label{ap:a1_5}

\begin{figure}
    \centering
    \includegraphics[width=2.5in]{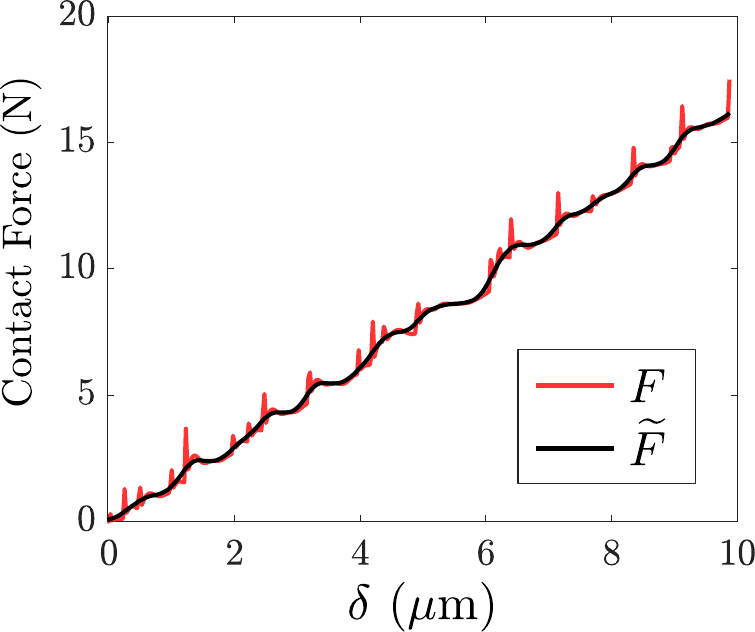}
    \caption{Raw and filtered indentation force $F, \widetilde F$ vs indentation depth $\delta$ for a typical dynamic indentation simulation.}
    \label{fig:filter_demo}
\end{figure}

One of the quantities of interest is the total reaction force on the indenter, $F(t) \coloneq \int_{\partial_C \Omega} -\lambda \pdv{\mathcal{C}}{u} \cdot e_3 dS $.  However, the explicit update of the displacements and contact introduces small spikes in the net load as shown in Figure~\ref{fig:filter_demo}.  We filter these away through mollification using a kernel in time and consider the force $\widetilde F  = K*F$ where $K \in H^1((-T, T)), \ K \ge 0$ with compact support in an interval $(-t_f,t_f),\  t_f  \ll T$ and $\int_{-t_f}^{t_f} K dt = 1$. For this work, we consider a piecewise linear ``hat" function
\begin{equation}
    K(t) = \begin{cases}
        \frac{1}{t_f^2} (t_f + t) & \text{ for } -t_f \leq t < 0, \\
        \frac{1}{t_f^2} (t_f - t)  & \text{ for } 0 \leq t \leq t_f, \\
        0 & \text{ otherwise }.
    \end{cases}
\end{equation} Figure~\ref{fig:filter_demo} compares the unfiltered and filtered contact forces.

\section{Adjoint method} \label{ap:a2}
We derive the adjoint relations for the dynamic rigid contact problem with viscoplastic evolution. For simplicity, we consider an objective of integral from 
\begin{equation}
    \mathcal{O}(u, q, \varepsilon^p, \lambda, P) = \int_0^T   \int_{\Omega} o(u, q, \varepsilon^p, \lambda, P) \, d\Omega.
\end{equation}
However, this derivation may be easily extended more complex objective functions. We seek an expression for the sensitivity of $\mathcal{O}$ with respect to the parameters $P$. Following the adjoint method, we augment the objective using the governing equations from~\eqref{eq:contact_evo},
    \begin{equation}
    \begin{aligned}
    \mathcal{O} = &\int_0^T  \int_{\Omega} \bigg\{ o + \rho \ddot{u} \cdot v + \mathbb{C}\varepsilon^e \cdot  \nabla v - b \cdot v + \gamma \dot{q} \left[ \sigma_M - \pdv{W^p}{q} - \pdv{{g}^*}{\dot{q}} \right] + \zeta \cdot \left( \dot{\varepsilon}^p - \dot{q} M \right) \bigg\} \, d\Omega  dt \\
    & - \int_0^T \int_{\partial_C \Omega} \lambda \pdv{\mathcal{C}_I}{u} \cdot v \ dS + \int_0^T \int_{\partial_C \Omega} \tau \lambda \left( \mathcal{C}_I - \ell^2 \right) \ dS dt.
    \end{aligned}
    \end{equation}
 where the fields $v$, $\gamma$, $\zeta$ and $\tau$  which correspond to the displacement, plastic hardening, plastic strain and Lagrange multiplier for contact respectively are to be determined.   Also, we have used the Kuhn-Tucker condition to replace (\ref{eq:contact_evo})$_2$ to include the irreversibility of the accumulated plastic strain.  We differentiate this augmented objective with respect to the parameters $P$,
 \begin{equation} \label{eq:long_adj}
    \begin{aligned}
    	\dv{\mathcal{O}}{P} &= \int_{0}^T  \int_{\Omega} \bigg\{ \pdv{o}{P} + \pdv{\mathbb{C}}{P} \varepsilon^e \cdot \nabla v  + \gamma \dot{q} \left( \pdv{{\sigma}_M}{P} - \pdv{^2 W^p}{q \partial P} - \pdv{^2 {g}^*}{\dot{q} \partial P}\right)  \\
    	& \quad + \pdv{o}{u} \cdot \dv{u}{P} + \rho v  \cdot \dv{\ddot{u}}{P} + \left( \mathbb{C} \nabla v  +  \gamma \dot{q} \pdv{{\sigma}_M}{\varepsilon} - \dot{q} \zeta \cdot \pdv{M}{\varepsilon} \right) \cdot \nabla \dv{u}{P}  \\
    	& \quad + \left( \pdv{o}{q}  - \gamma \dot{q} \pdv{^2W^p}{q^2}   \right) \dv{q}{P}+   \left( \gamma  \left[ \sigma_M - \pdv{W^p}{q} - \pdv{{g}^*}{\dot{q}} \right] - \gamma \dot{q} \pdv{^2{g}^*}{\dot{q}^2}  - \zeta \cdot M  \right) \dv{\dot{q}}{P} \\
    	& \quad + \zeta \cdot \dv{\dot{\varepsilon}^p}{P}  + \left( \pdv{o}{\varepsilon^p} - \nabla v \cdot  \mathbb{C}  + \gamma \dot{q}  \pdv{{\sigma}_M}{\varepsilon^p} - \dot{q} \zeta \cdot \pdv{M}{\varepsilon^p} \right) \cdot  \dv{\varepsilon^p}{P}  \bigg \} \ d\Omega \ dt \\ 
        & \quad + \int_0^T \int_{\partial_C \Omega} \left(\tau \lambda \pdv{\mathcal{C}_I}{u} -\lambda  v \cdot \pdv{^2\mathcal{C}_I}{u \partial u}  \right) \cdot \dv{u}{P} \ dS \ dt  \\
        & \quad + \int_0^T \int_{\partial_C \Omega_{\lambda \neq 0}} \left(\pdv{\mathcal{C}_I}{u} \cdot v -\pdv{o}{\lambda}  \right) \cdot \dv{\lambda}{P} \ dS \ dt .
    \end{aligned}
    \end{equation}
Following which, we integrate by parts, and enforce quiescent conditions on $v$ at the final time $T$ to remove boundary terms. This leads to the adjoint problem 
\begin{equation}
    \begin{aligned}
    &0 =\int_{\Omega} \left[ \rho \ddot{v} \cdot \delta u +  \left(  \mathbb{C} \nabla v +  \gamma \dot{q} \pdv{\sigma_M}{\varepsilon} - \dot{q} \zeta \cdot \pdv{M}{\varepsilon} \right) \cdot \nabla \delta u  + \pdv{o}{u} \cdot \delta u \right] \ d\Omega \\
    &\qquad + \int_{\partial_C \Omega} \left( \tau \lambda \pdv{\mathcal{C}_I}{u}   - \lambda v \cdot \pdv{^2 \mathcal{C}_I}{u \partial u} \right) \cdot \delta u \ dS   && \forall \delta u  \in \mathcal{K}_0, \\
    & \dv{}{t} \left[ \gamma \left( \sigma_M - \sigma_0 - \pdv{g^*}{\dot{q}}\right) - \gamma \dot{q} \pdv{^2 g^*}{\dot{q}^2} - \zeta \cdot M \right]  = \pdv{o}{q} - \gamma \dot{q}  \pdv{^2 W^p}{q^2} && \text{ on } \Omega, \\
    & \dv{\zeta}{t} =  \pdv{o}{\varepsilon^p}- \mathbb{C} \nabla v + \gamma \dot{q} \pdv{\sigma_M}{\varepsilon^p} - \dot{q} \zeta \cdot \pdv{M}{\varepsilon^p} && \text{ on } \Omega, \\
    & 0 = \pdv{\mathcal{C}_I}{u} \cdot v - \pdv{o}{\lambda} && \text{ on } \partial_C \Omega_{\lambda \neq 0}, \\
     & v|_{t = T} = \dot{v} |_{t = T} = 0, \ \gamma |_{t = T} = 0, \quad  \mu |_{t = T} = 0,
    \end{aligned}
\end{equation}
where $\mathcal{K}_0 \coloneq \{ \varphi \in H^1(\Omega), \ \varphi = 0 \text{ on } \partial_u \Omega \}$ is the space of kinematically admissible displacement variations. Then the sensitivity simplifies to 
\begin{equation}
    \dv{\mathcal{O}}{P} = \int_{0}^T  \int_{\Omega} \left[ \pdv{o}{P}  + \pdv{\mathbb{C}}{P} \varepsilon^e \cdot \nabla v  + \gamma \dot{q} \left( \pdv{{\sigma}_M}{P} - \pdv{^2 W^p}{q \partial P} - \pdv{^2 {g}^*}{\dot{q} \partial P}\right)  \right ] \ d\Omega \ dt.
\end{equation}

{The above derivation neglects the variation of the contact area $\partial_C \Omega_{\lambda \neq 0}$ with $u$ and $\lambda$. That is, we neglect $\dv{\lambda}{P}$ in regions where $\lambda = 0$. This is only invalid for special cases, that is, when the contact force is discontinuous on the boundary of the contact region. Further, the unilateral constraint renders derivatives in these cases ill-defined. However, once the system is discretized, these issues are no longer present. The discrete problem has ill-defined derivatives only when nodes are exactly at the onset of contact, and this occupies a measure zero set in state-space. Thus, the method is fully accurate in practical settings. }

\section{Numerical method for the adjoint problem} \label{ap:a3}
We detail the numerical scheme used to solve the adjoint relations. The boundary conditions for the adjoint relations in ~\eqref{eq:adj_rels} are specified at the final time $T$. Thus, we solve this system backwards in time. The adjoint variable set $\{v,\ \gamma,\ \zeta, \ \tau \}$ takes the same spacial discretization as their forward variable counterparts $\{u, \ q, \ \varepsilon^p, \ \lambda \}$, and we update them in a similar manner. 

 We compute a predictor for the adjoint displacement and velocities explicitly while ignoring the adjoint contact relation. Then, we compute the adjoint contact force $\tau$ on the nodes which were in contact in the forward problem. This is used to correct the adjoint displacements and velocities. Finally, the adjoint plastic variables $\gamma$, $\zeta$ are updated implicitly.

 For the $n+1$ to $n$ timestep the predictor displacements are
\begin{equation}\label{eq:contact_adj}
    \begin{aligned}
        \ddot{v}_i^{n+1,\text{pre}} &= \frac{1}{m_i} H_i^{n+1}, \\
        \dot{v}_i^{n + 1/2, \text{pre}} &= \dot{v}_i^{n + 3/2} - \Delta t^{n+1} \, \ddot{v}_i^{n +1, \text{pre}}, \\
        v_i^{n, \text{pre}} &= v_i^{n+1} - \Delta t^{n + 1/2} \, \dot{v}_i^{n + 1/2},
    \end{aligned}
\end{equation}
where
\begin{equation}
\begin{aligned}
    H_i^n &= \int_{\Omega} \left[ - \left( \nabla v^n \cdot \mathbb{C}  +  \gamma^n \dot{q}^n \pdv{{\sigma}_M}{\varepsilon} - \dot{q}^n \zeta^n \cdot \pdv{M}{\varepsilon}\right) \cdot \nabla N_i  - \frac{\partial o}{\partial u} \cdot N_i \right] \, d\Omega  \\
    &\qquad + \int_{\partial_C \Omega} \left(\lambda^n v^n \cdot \pdv{^2 \mathcal{C}_I}{u \partial u} \right) \cdot N_i \ dS. \\
\end{aligned}
\end{equation}
Next, we compute a correction force to satisfy the adjoint contact constraint. For the nodes on $\partial_C \Omega$ at which $\lambda \neq 0$ we compute the adjoint corrective force
\begin{equation}
    H^c_i = \frac{m_i}{\Delta t^{n+1} \Delta t^{n + 1/2}} \left( \pdv{o}{\lambda_i}  - v^{n,\text{pre}}_i \cdot e_r \right) e_r,
\end{equation}
where $e_r$ is the outward unit normal of the indenter. The adjoint displacements and velocities are corrected by
\begin{equation}
\begin{aligned}
    \dot{v}_i^{n + 1/2} &= \dot{v}_i^{n + 1/2, \text{pre}} - \Delta t^{n+1} \frac{H^c_i}{m_i},\\
    {v}_i^{n} &= {v}_i^{n, \text{pre}} + \Delta t^{n + 1/2} \Delta t^{n+1} \frac{H^c_i}{m_i}.
    \end{aligned}
\end{equation}
We then update the adjoint variables $\gamma$ and $\zeta$ through an implicit forward Euler discretization. For this, we solve through direct inversion a linear set of equations point-wise at each quadrature point
\begin{equation}
\begin{aligned}
&\frac{1}{\Delta t^{n+1/2}} \left[ \gamma \left( {\sigma}_M - \pdv{W^p}{q} - \pdv{{g}^*}{\dot{q}} -  \dot{q} \pdv{^2 {g}^*}{\dot{q}^2}\right)
- \zeta \cdot M \right]_{t^{n}}^{t^{n+1}} = \left. \pdv{o}{q} \right |_{t_n} - \gamma^n \left( \dot{q} \pdv{^2W^p}{q^2} \right)_{t_n} \\
& \frac{\zeta^{n+1} - \zeta^n}{\Delta t^{n+1/2}} =  \left. \pdv{o}{\varepsilon^p} \right |_{t_n} + \nabla v^n \cdot \left. \pdv{^2 W^e}{\varepsilon \partial \varepsilon^p} \right |_{t_n} + \gamma ^n \left( \dot{q} \pdv{{\sigma}_M}{\varepsilon^p} \right)_{t_n} - \zeta^n \cdot \left( \dot{q} \pdv{M}{\varepsilon^p}\right)_{t_n}.
\end{aligned}
\end{equation}

{
\section{Modification for Ellipsoidal Indenters} \label{ap:a4}

We highlight the modifications to the formulation and numerical scheme to account for an ellipsoidal indenter. We consider an ellipsoidal indenter with semi-axes $\{R_x, R_y, R_z \}$ with contact function
\begin{equation}
    \mathcal{C}(X + u, x_I(t)) \coloneqq \frac{ 
    \left( (X + u - x_I) \cdot e_1 \right )^2}{R_x^2} + \frac{ 
    \left( (X + u - x_I) \cdot e_2 \right )^2}{R_y^2} + \frac{ 
    \left( (X + u - x_I) \cdot e_3 \right )^2}{R_z^2} - 1.
\end{equation}
Compared to the spherical case, this has two key differences. The first is that the interpenetration distance can no longer be expressed as a simple expression, and we most solve a non-linear equation for this quantity. The second difference is that the derivative of the contact function $
\pdv{\mathcal{C}}{u}$ does not have unit magnitude. These differences must be propagated through the numerical scheme for both the forward and adjoint problem, and we describe the details of this in the following.

\paragraph*{Forward Problem.} The only difference from the spherical case appears in the interpenetration distance and contact force expression. If contact is detected, we solve
\begin{equation}
    0 = \mathcal{C} \left( X_i + u_i^{n+1, \text{pre}} + \delta_i \norm{\pdv{\mathcal{C}}{u}}^{-1}\pdv{\mathcal{C}}{u},\  x_I \right). 
\end{equation}
This corresponds to finding the closest point on the ellipsoid surface from the predictor position. The distance between these is $\delta_i$. We solve this through a Newton-Raphson scheme, capping the number of iterations at five. 

The contact force is computed as
\begin{equation}
     F^c_i = \frac{\delta_i m_i}{\Delta t^n \Delta t^{n + 1/2}} \norm{\pdv{\mathcal{C}}{u}}^{-1}\pdv{\mathcal{C}}{u},
\end{equation}
and the Lagrange multiplier is simply
\begin{equation}
    \lambda_i = -\norm{\pdv{\mathcal{C}}{u}}^{-1} \norm{F^c_i}.
\end{equation}

\paragraph{Adjoint Problem.} Because the gradient of the contact function no longer has unit magnitude, the adjoint corrective force takes a modified form. Following Appendix~\ref{ap:a3}, the adjoint corrective force becomes
\begin{equation}
    H^c_i = \frac{m_i}{\Delta t^{n+1} \Delta t^{n + 1/2}} \left( \pdv{o}{\lambda_i}  - v^{n,\text{pre}}_i \cdot \pdv{\mathcal{C}}{u} \right) \norm{\pdv{C}{u}}^{-2} \pdv{\mathcal{C}}{u},
\end{equation}
with the rest of the numerical scheme remaining identical.
}

{
\section{AL 6061-T6 Indentation Comparison} \label{ap:a5}

We compare the indentation response of AL 6061-T6 using the recovered model to the literature Johnson Cook model from Clayton et. al~\cite{clayton2023simulation}, both using rigid spherical indenters. This is shown in Figure~\ref{fig:AL_JC_compare}. We observe that the recovered model is a closer match to the experimental data than the Johnson Cook model, and note that the objective function computed for the recovered model ($\mathcal{O} = 2.94 \times 10^{-4}$) is an order of magnitude lower than that of the Johnson Cook ($\mathcal{O} = 2.24 \times 10^{-3}$). This result is not surprising, as the recovered model was fit to match this exact experimental curve. However, it highlights the degree of accuracy that the fit achieves.
\begin{figure}[htb!]
    \centering
    \includegraphics[width=0.6\linewidth]{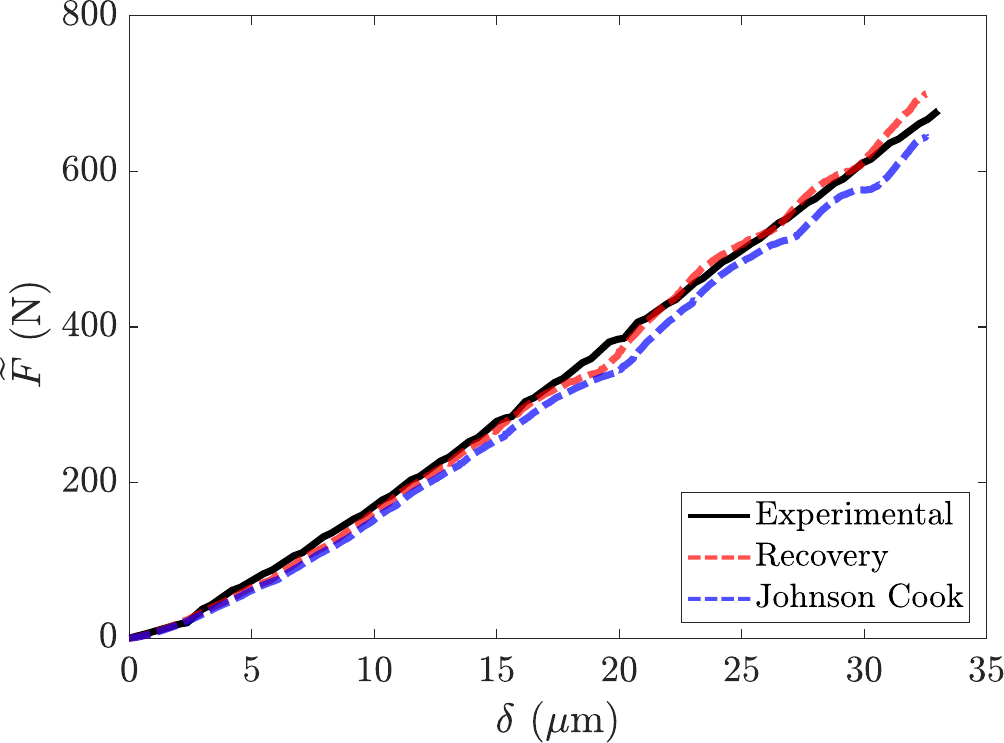}
    \caption{Comparison of the recovered force vs displacement for indentation on AL 6061-T6, showing the experimental data (black), simulations using the recovered model (red dashes), and simulations using the Johnson Cook model from Clayton et al~\cite{clayton2023simulation} with parameters shown in Table~\ref{tab:JC_params}.}
    \label{fig:AL_JC_compare}
\end{figure}
}

\end{document}